\documentclass{ws-ijmpcs}

\usepackage{amsmath}
\usepackage{latexsym}
\usepackage{amssymb}
\usepackage{dsfont}

\newcommand{\ud}{\mathrm{d}}

\newcommand{\uTr}{\mathrm{Tr}}

\newcommand{\uvec}[1]{\boldsymbol{#1}}
\newcommand{\pure}{\text{pure}}
\newcommand{\phys}{\text{phys}}

\begin{document}

\markboth{C. Lorc\'e}
{The proton spin decomposition : path dependence and gauge symmetry}

%
\catchline{}{}{}{}{}
%

\title{THE PROTON SPIN DECOMPOSITION : PATH DEPENDENCE AND GAUGE SYMMETRY}

\author{C\'EDRIC LORC\'E}

\address{IPNO, Universit\'e Paris-Sud, CNRS/IN2P3, 91406 Orsay, France\\
and LPT, Universit\'e Paris-Sud, CNRS, 91405 Orsay, France\\
lorce@ipno.in2p3.fr}

\maketitle


\begin{abstract}
We discuss the different types of decomposition of the proton spin. We stress that, despite their lack of uniqueness, the Chen \emph{et al.} and Wakamatsu decompositions are perfectly measurable. We argue that a large part of the recent controversies boils down to the fact that there actually exist two types of gauge transformations in the Chen \emph{et al.} approach, where physical and gauge degrees of freedom of the gauge potential are explicitly separated. By carefully distinguishing these two types of gauge transformations, one can easily understand how the concepts of gauge invariance, Stueckelberg symmetry, path dependence and measurability are linked to each other.
\end{abstract}


\section{Introduction}	

The problem of decomposing the proton spin into quark/gluon and spin/orbital angular momentum (OAM) contributions has been revived a few years ago by Chen \emph{et al.}\cite{Chen:2008ag,Chen:2009mr}. By separating explicitly the gauge degrees of freedom from the physical ones, they challenged the textbook knowledge\cite{Jauch,Berestetskii,Jaffe:1989jz,Ji:1996ek} in providing a complete gauge-invariant decomposition. This approach triggered many new developments and reopened old controversies regarding the measurability and the physical interpretation of the different contributions.

In this proceeding, we summarize the situation and argue that a large part of the controversies can in fact be resolved once the concepts and requirements are made clear. In section \ref{sec2}, we briefly present the four main different types of proton spin decompositions. In particular, we show for the first time that the gluon kinetic/Wakamatsu OAM can be put in a form ``$\uvec r\times\uvec p$''. In section \ref{sec3}, we review the approach proposed by Chen \emph{et al.} and discuss the issue of its non-uniqueness resulting from the Stueckelberg symmetry of the original Lagrangian. We show in particular in section \ref{sec4} that this Stueckelberg symmetry simply originates from the possibility of treating differently active and passive gauge transformations in the Chen \emph{et al.} approach, owing to the presence of background fields like in General Relativity. In section \ref{sec5} we argue that measurable quantities are not necessarily invariant under Stueckelberg transformations. Indeed, the Chen \emph{et al.} approach turns out to be simple procedure for constructing non-local gauge-invariant extensions of gauge non-invariant quantities. The Stueckelberg dependence can therefore be interpreted as a generalization of the path dependence of non-local gauge-invariant quantities. Following these observations, we propose in section \ref{sec6} to sort measurable quantities into two categories according to their behavior under Stueckelberg transformations, allowing one to reconcile contradictory claims in the literature. In particular, the status of the different proton spin decompositions is commented. Finally, we conclude with section \ref{sec7}.

\section{The decompositions in a nutshell}\label{sec2}

There exist essentially four types of proton spin decompositions into quark/gluon and spin/OAM contributions\cite{Lorce:2012rr}. We refer to them as the Jaffe-Manohar\cite{Jaffe:1989jz}, Ji\cite{Ji:1996ek}, Chen \emph{et al.}\cite{Chen:2008ag,Chen:2009mr} and Wakamatsu\cite{Wakamatsu:2010qj,Wakamatsu:2010cb} decompositions. They are respectively given by
\begin{align}
\uvec J_\text{QCD}&=\uvec S^q_\text{JM}+\uvec L^q_\text{JM}+\uvec S^g_\text{JM}+\uvec L^g_\text{JM},\\
&=\uvec S^q_\text{Ji}+\uvec L^q_\text{Ji}+\uvec J^g_\text{Ji},\\
&=\uvec S^q_\text{Chen}+\uvec L^q_\text{Chen}+\uvec S^g_\text{Chen}+\uvec L^g_\text{Chen},\\
&=\uvec S^q_\text{Wak}+\uvec L^q_\text{Wak}+\uvec S^g_\text{Wak}+\uvec L^g_\text{Wak}.
\end{align}

In the quark sector, all the decompositions agree on the spin contribution
\begin{equation}
\uvec S^q_\text{JM}=\uvec S^q_\text{Ji}=\uvec S^q_\text{Chen}=\uvec S^q_\text{Wak}=\int\ud^3x\,\psi^\dag\tfrac{1}{2}\uvec\Sigma\psi.
\end{equation}
but give different definitions for the OAM
\begin{align}
\uvec L^q_\text{JM}&=\int\ud^3x\,\psi^\dag(\uvec x\times\tfrac{1}{i}\uvec\nabla)\psi,\\
\uvec L^q_\text{Ji}&=\uvec L^q_\text{Wak}=\int\ud^3x\,\psi^\dag(\uvec x\times i\uvec D)\psi,\\
\uvec L^q_\text{Chen}&=\int\ud^3x\,\psi^\dag(\uvec x\times i\uvec D_\pure)\psi,
\end{align}
where the covariant derivative is given by $\uvec D=-\uvec\nabla-ig\uvec A$. In the Chen \emph{et al.} decomposition, one assumes that the gauge potential can be written as a sum of two terms $\uvec A=\uvec A_\pure+\uvec A_\phys$ (see next section) and defines accordingly a pure-gauge covariant derivative as $\uvec D_\pure=-\uvec\nabla-ig\uvec A_\pure$. 

In the gluon sector, the spin contributions read
\begin{align}
\uvec S^g_\text{JM}&=\int\ud^3x\,\uvec E^a\times\uvec A^a,\\
\uvec S^g_\text{Chen}&=\uvec S^g_\text{Wak}=\int\ud^3x\,\uvec E^a\times\uvec A^a_\phys,
\end{align}
where $a$ is an adjoint color index $\uvec A=\uvec A^at^a$ with $2\uTr[t^at^b]=\delta^{ab}$, and the OAM contributions read
\begin{align}
\uvec L^g_\text{JM}&=\int\ud^3x\,E^{ai}(\uvec x\times\uvec\nabla) A^{ai},\\
\uvec L^g_\text{Chen}&=-\int\ud^3x\,E^{ai}(\uvec x\times\uvec{\mathcal D}^{ab}_\pure) A^{bi}_\phys,\\
\uvec L^g_\text{Wak}&=\uvec L^g_\text{Chen}-\int\ud^3x\,(\uvec{\mathcal D}\cdot\uvec E)^a\,\uvec x\times\uvec A^a_\phys,
\end{align}
where the ordinary and pure-gauge covariant derivatives in the adjoint representation are given by $\uvec{\mathcal D}^{ab}=-\uvec\nabla\,\delta^{ab}-gf^{abc}A^c$ and $\uvec{\mathcal D}^{ab}_\pure=-\uvec\nabla\,\delta^{ab}-gf^{abc}A^c_\pure$, respectively, with $[t^a,t^b]=if^{abc}t^c$. 

As one can easily see, the Chen \emph{et al.} and Wakamatsu decompositions simply differ in the attribution of the potential OAM to either quarks or gluons
\begin{equation}
\uvec L_\text{pot}=-\int\ud^3x\,(\uvec{\mathcal D}\cdot\uvec E)^a\,\uvec x\times\uvec A^a_\phys=\uvec L^g_\text{Wak}-\uvec L^g_\text{Chen}=\uvec L^q_\text{Chen}-\uvec L^q_\text{Wak},
\end{equation}
where the QCD equation of motion $(\uvec{\mathcal D}\cdot\uvec E)^a=g\psi^\dag t^a\psi$ has been used in the last equality. The only difference between the Ji and Wakamatsu decompositions is that the Ji decomposition does not provide any splitting of the total gluon angular momentum into spin and OAM contributions
\begin{equation}
\uvec J^g_\text{Ji}=\uvec S^g_\text{Wak}+\uvec L^g_\text{Wak}=\int\ud^3x\,\uvec x\times(\uvec E^a\times\uvec B^a).
\end{equation}
Dropping a surface term, we find the following compact expression for the gluon OAM in the Wakamatsu decomposition
\begin{equation}
\uvec L^g_\text{Wak}=\int\ud^3x\,\uvec x\times[(\uvec A_\phys\times\uvec{\mathfrak D}_\pure)^a\times\uvec E^a],
\end{equation}
where $\uvec{\mathfrak D}_\pure=\tfrac{1}{2}(\uvec{\mathcal D}+\uvec{\mathcal D}_\pure)$. This expression has the form of an OAM term since the kinetic gluon momentum can be written as
\begin{equation}
\uvec P^g_\text{kin}=\int\ud^3x\,\uvec E^a\times\uvec B^a=\int\ud^3x\,(\uvec A_\phys\times\uvec{\mathfrak D}_\pure)^a\times\uvec E^a,
\end{equation}
where a surface term has also been dropped.

\section{The Chen \emph{et al.} approach}\label{sec3}

Both the Chen \emph{et al.} and Wakamatsu decompositions make use of a splitting of the gauge potential into \emph{pure-gauge} and \emph{physical} terms\cite{Chen:2008ag,Chen:2009mr,Lorce:2012rr,Wakamatsu:2010qj,Wakamatsu:2010cb}
\begin{equation}\label{decomposition}
A_\mu(x)=A^\pure_\mu(x)+A^\phys_\mu(x),
\end{equation}
where, by definition, the pure-gauge term does not contribute to the field strength
\begin{equation}\label{cond1}
F^\pure_{\mu\nu}=\partial_\mu A^\pure_\nu-\partial_\nu A^\pure_\mu-ig[A^\pure_\mu,A^\pure_\nu]=0
\end{equation}
and transforms as
\begin{equation}\label{cond2}
A^\pure_\mu(x)\mapsto \tilde A^\pure_\mu(x)=U(x)[A^\pure_\mu(x)+\tfrac{i}{g}\partial_\mu]U^{-1}(x)
\end{equation}
under gauge transformations. As a consequence $A^\pure_\mu(x)$ can be written as a pure-gauge term
\begin{equation}
A^\pure_\mu(x)=\tfrac{i}{g}\,U_\pure(x)\partial_\mu U^{-1}_\pure(x),
\end{equation}
where $U_\pure(x)$ is some unitary matrix with the gauge transformation law
\begin{equation}\label{gaugeT}
U_\pure(x)\mapsto\tilde U_\pure(x)=U(x)U_\pure(x).
\end{equation}
The physical term is responsible for the field strength
\begin{equation}
F_{\mu\nu}=\mathfrak D^\pure_\mu A^\phys_\nu-\mathfrak D^\pure_\nu A^\phys_\mu,
\end{equation}
and transforms like the latter under gauge transformations
\begin{equation}
A^\phys_\mu(x)\mapsto \tilde A^\phys_\mu(x)=U(x)A^\phys_\mu(x)U^{-1}(x).
\end{equation}
Interestingly, the matrix $U_\pure(x)$ can be interpreted as a dressing field and used to define gauge-invariant fields\cite{Chen:2012vg,Lorce:2013gxa} known as Dirac variables\cite{Dirac:1955uv,DeWitt:1962mg,Mandelstam:1962mi}
\begin{align}
\hat\psi(x)&\equiv U^{-1}_\text{pure}(x)\psi(x),\\
\hat A_\mu(x)&\equiv U^{-1}_\text{pure}(x)[A_\mu(x)+\tfrac{i}{g}\,\partial_\mu]U_\text{pure}(x)
\end{align}
with $\hat A_\mu(x)=\hat A^\phys_\mu(x)$ and $\hat A^\pure_\mu(x)=0$.

The main problem with the Chen \emph{et al.} approach is that the splitting of the gauge potential is not unique. Indeed, starting from a given splitting \eqref{decomposition}, one can easily construct a new splitting $A_\mu(x)=\bar A^\pure_\mu(x)+\bar A^\phys_\mu(x)$ satisfying the conditions \eqref{cond1} and \eqref{cond2} as follows
\begin{equation}
\begin{aligned}
\bar A^\pure_\mu(x)&=A^\pure_\mu(x)+B_\mu(x),\\
\bar A^\phys_\mu(x)&=A^\phys_\mu(x)-B_\mu(x),
\end{aligned}
\end{equation}
where $B_\mu(x)=\frac{i}{g}\,U_\pure(x)U^{-1}_S(x)[\partial_\mu U_S(x)]U^{-1}_\pure(x)$ with $U_S(x)$ a gauge-invariant matrix. Because of the similarity between the Chen \emph{et al.} approach and the Stueckelberg mechanism, the transformation $\phi(x)\mapsto\bar\phi(x)$ is referred to as the Stueckelberg transformation\cite{Lorce:2012rr,Stoilov:2010pv}. Note also that while gauge transformations act on the left of $U_\pure(x)$, see Eq. \eqref{gaugeT}, Stueckelberg transformations act on the right
\begin{equation}\label{Stueck}
\bar U_\pure(x)=U_\pure(x)U^{-1}_S(x).
\end{equation}
In this approach, the physical term is treated as a dynamical variable, while the pure-gauge term plays the role of an external, auxiliary, non-dynamical variable. One can then interpret $A^\pure_\mu(x)$ and $U_\pure(x)$ as \emph{background} fields\cite{Lorce:2013gxa,Zhang:2011rn}. Stueckelberg dependence is therefore simply background dependence.

\section{Active and passive gauge transformations}\label{sec4}

By construction, the Chen \emph{et al.} and Wakamatsu decompositions are formally gauge invariant albeit not unique, since they are obviously not invariant under Stueckelberg transformations. This lack of uniqueness is often considered as a signature that the Chen \emph{et al.} approach is not really gauge invariant in the textbook sense. Such a statement is however pretty confusing and requires some clarification\cite{Wakamatsu:2013voa}.

Usually, one does not specify whether gauge transformations have to be considered as \emph{active} or \emph{passive} transformations. The reason is that such a distinction is impossible to make as long as one deals with dynamical fields only. One is free to adopt her favorite point of view. Particle physicists often consider gauge symmetry as a mere redundancy of the mathematical formalism, and so gauge transformations are implicitly thought of as passive transformations. There is indeed in principle no way to experimentally fix or determine a gauge. On the other hand, mathematicians usually think of gauge symmetry as a property of the system, and accordingly consider gauge transformations as active transformations.

The situation changes however once a background field is introduced into the game\cite{Smolin:2005mq,Rickles,Rozali:2008ex,Barenz:2012av}. Passive gauge transformations simply change the coordinate axes in internal space, and therefore affect in the same way dynamical and background fields. Active gauge transformations change the fields themselves, but treat differently dynamical and background fields.

Since in the Chen \emph{et al.} approach the background field $U_\pure(x)$ is assumed to transform in the same way as the fermion field $\psi(x)$, the gauge transformations considered by Chen \emph{et al.}
\begin{align}
A_\mu(x)&\mapsto\tilde A_\mu(x)=U_p(x)[A_\mu(x)+\tfrac{i}{g}\partial_\mu]U^{-1}_p(x),\\
\psi(x)&\mapsto\tilde\psi(x)=U_p(x)\psi(x),\\
U_\pure(x)&\mapsto\tilde U_\pure(x)=U_p(x)U_\pure(x),\label{passive2}
\end{align}
correspond actually to passive gauge transformations. To stress this, we have added an index $p$ to the gauge transformation matrix $U(x)$. We observed that one could have considered instead a different type of gauge transformations\cite{Lorce:2013bja}
\begin{align}
A_\mu(x)&\mapsto\check A_\mu(x)=U_a(x)[A_\mu(x)+\tfrac{i}{g}\partial_\mu]U^{-1}_a(x),\\
\psi(x)&\mapsto\check\psi(x)=U_a(x)\psi(x),\\
U_\pure(x)&\mapsto\check U_\pure(x)=U_a(x)U_\pure(x)U^{-1}_a(x).
\end{align}
Clearly, these gauge transformations treat differently the fermion field $\psi(x)$ and the background field $U_\pure(x)$, and correspond therefore to active gauge transformations. To stress this, we have added an index $a$ to the gauge transformation matrix $U(x)$. 

Now, combining an active gauge transformation with the corresponding inverse passive gauge transformation $U_a(x)=U^{-1}_p(x)\equiv U_S(x)$, we see that the gauge field $A_\mu(x)$ and the fermion field $\psi(x)$ remain unchanged, while the background field $U_\pure(x)$ is changed
\begin{align}
A_\mu(x)&\mapsto A_\mu(x),\\
\psi(x)&\mapsto\psi(x),\\
U_\pure(x)&\mapsto U_\pure(x)U^{-1}_S(x),
\end{align}
which coincides with Eq. \eqref{Stueck}. We therefore see that Stueckelberg symmetry simply reflects our freedom in treating differently passive and active gauge transformations in presence of a background field. This clarifies how gauge and Stueckelberg symmetries are related to each other.

In order to avoid any confusion in the following discussions, instead of working with the set of active and passive gauge transformations, we work with the equivalent set of Stueckelberg and (passive) gauge transformations.

\section{Gauge symmetry, measurability and path dependence}\label{sec5}

We have seen that the Chen \emph{et al.} and Wakamatsu decompositions are not unique. In practice, one explicitly breaks the Stueckelberg symmetry by imposing an additional condition on the physical term. For example, Chen \emph{et al.}\cite{Chen:2008ag,Chen:2009mr} imposed the Coulomb constraint $\uvec\nabla\cdot\uvec A_\phys=0$ motivated by the famous Helmoltz decomposition of QED, while Hatta\cite{Hatta:2011zs,Hatta:2011ku} imposed instead the light-front constraint $A^+_\phys=0$ with the aim of making contact with the parton model of QCD. Contrary to \emph{e.g.} Chen \emph{et al.} and Wakamatsu, our point of view is that there is \emph{a priori} no fundamental argument which determines what Stueckelberg-fixing constraint to use. Such a choice is only dictated by reasons of convenience, just like the choice of a gauge.

Since the gauge symmetry is not physical, measurable quantities are necessarily gauge invariant and gauge non-invariant quantities are not measurable. However, it is perfectly possible that a gauge non-invariant quantity evaluated in a certain gauge $G$ gives formally the same answer as some gauge-invariant quantity. The latter can then be considered as a \emph{gauge-invariant extension} (GIE) of the former\cite{Hoodbhoy:1998bt,Ji:2012sj}. Strictly speaking, one does not measure gauge non-invariant quantities in a fixed gauge, but only their GIEs. The GIE is usually a complicated non-local expression with \emph{a priori} no simple gauge-invariant interpretation. Only in the gauge $G$ is the interpretation simple because of the formal equivalence with a local gauge non-invariant expression. 

The archetypical example is the quantity\cite{Manohar:1990kr} $\Delta g=\int\ud x\,[g_+(x)-g_-(x)]$, where $g_\lambda(x)$ is the PDF of gluons with polarization $\lambda=\pm 1$. Such a quantity is measurable and has a complicated non-local gauge-invariant expression, but formally coincides with the expectation value of the Jaffe-Manohar gluon helicity operator $S^{g,z}_\text{JM}$ in the light-front gauge. In other words, $\Delta g$ is the light-front GIE of the expectation value of $S^{g,z}_\text{JM}$. This is what is actually meant when one carelessly says that the gluon spin is measurable. 

In the Chen \emph{et al.} approach, one can always find a gauge such that the pure-gauge term vanishes $A_\mu(x)=A^\phys_\mu(x)$: it suffices to consider the gauge transformation with $U(x)=U^{-1}_\pure(x)$. The Chen \emph{et al.} decomposition can therefore be seen as a GIE of the Jaffe-Manohar decomposition\cite{Ji:2012sj}. In particular, like the Jaffe-Manohar operators, the Chen \emph{et al.} operators satisfy the canonical commutation relations and are generators of Poincar\'e transformations. This boils down to the fact that the Chen \emph{et al.} approach allows one to reconcile the canonical formalism with gauge symmetry\cite{Lorce:2013gxa,Guo:2012wv,Guo:2013jia}. Strictly speaking, the Jaffe-Manohar decomposition is not measurable because it is gauge non-invariant. But seen as a fixed-gauge decomposition, it can be considered to be measurable simply because there exists a corresponding GIE in the form of a Chen \emph{et al.} decomposition.

As mentioned earlier, GIEs have usually complicated non-local expressions. This non-locality is simply hidden in the Chen \emph{et al.} notation. Indeed, in many cases the matrix $U_\pure(x)$ consists in a simple Wilson line $\mathcal W_\mathcal C(x,x_0)=\mathcal P[e^{ig\int^x_{x_0}\ud s^\mu A_\mu(s)}]$, where $x_0$ is a fixed reference point and $\mathcal C$ is a path parametrized by $s^\mu(t)$. The pure-gauge and physical terms can then be expressed as non-local functionals of the gauge potential\cite{Hatta:2011ku,Lorce:2012ce}
\begin{align}
A^\pure_\mu(x)&=\frac{i}{g}\,\mathcal W_\mathcal C(x,x_0)\frac{\partial}{\partial x^\mu}\mathcal W_\mathcal C(x_0,x),\\
A^\phys_\mu(x)&=-\int^x_{x_0}\mathcal W_\mathcal C(x,s)F_{\alpha\beta}(s)\mathcal W_\mathcal C(s,x)\,\frac{\partial s^\alpha}{\partial x^\mu}\,\ud s^\beta.
\end{align}
By merely changing the path $\mathcal C$, one changes the individual pure-gauge and physical terms but not their sum. The Stueckelberg/background dependence can therefore be understood as a kind of generalized path dependence.

\section{Observables and quasi-observables}\label{sec6}

It follows from the discussions above that the measurable gauge-invariant quantities can be sorted into two categories: the \emph{observables} which are Stueckelberg/background/path independent, and the \emph{quasi-observables} which are Stueckelberg/background/path dependent. Observables are gauge invariant in a strong sense, \emph{i.e.} under both passive and active transformations, which is what is usually meant by ``gauge invariant in the textbooks sense''. Quasi-observables are gauge invariant in a weak sense, \emph{i.e.} under passive transformations only, which is sufficient to be measurable.

Observables can be written in terms of local gauge-invariant operators and can in principle be accessed experimentally without the need of a kinematical expansion or factorization framework. On the contrary, quasi-observables cannot be written in terms of operators that are at the same time local and gauge invariant. The corresponding gauge-invariant operators are necessarily non-local while the corresponding local operators are necessarily gauge non-invariant. They can only be accessed experimentally provided that a suitable kinematical expansion or factorization framework is available. It is indeed the latter that breaks the Stueckelberg symmetry and determines the shape of the Wilson line to use. In QCD for example, the proton internal structure is probed in high-energy experiments. Factorization theorems indicate that Wilson lines essentially run along the light-front direction\cite{Collins}, so that the natural Stueckelberg fixing constraint is $A^+_\phys=0$.

Cross-sections, charges and form factors are clear examples of observables while parton distribution functions, which are path/scale/scheme-dependent, are typical examples of quasi-observables. One might argue that observables are the only truly physical quantities, but in our opinion this debate goes somewhat beyond physics considerations and is not essential regarding measurability. Note also that some observables can be obtained from quasi-observables. This is the essence of \emph{e.g.} the Ji relation\cite{Ji:1996ek} that expresses the energy-momentum form factors $A(0)$ and $B(0)$ in terms of particular moments of generalized parton distributions $H(x,\xi,t)$ and $E(x,\xi,t)$
\begin{equation}
J^{q,g}_\text{Ji}=\frac{1}{2}\left[A^{q,g}(0)+B^{q,g}(0)\right]=\frac{1}{2}\int\ud x\,x\left[H^{q,g}(x,0,0)+E^{q,g}(x,0,0)\right].
\end{equation}

According to textbooks\cite{Jauch,Berestetskii}, there cannot be any local gauge-invariant decomposition of the gluon angular momentum into spin and OAM contributions. It does not mean that they do not exist, but simply that they correspond to quasi-observables. In the Ji decomposition, only observables are allowed. It follows that this decomposition is unique, but fails in providing us with a decomposition of the gluon angular momentum into spin and OAM contributions. Since measurability is the only relevant criterion, one can in fact allow quasi-observables to be involved. Quasi-observables can be expressed either in terms of local fixed-gauge operators like in the Jaffe-Manohar decomposition, or as non-local gauge-invariant operators like in the Chen \emph{et al.} and Wakamatsu decompositions.

\section{Conclusion}\label{sec7}

We presented and discussed the four main types of proton spin decompositions. As many contradictory claims are spread in the literature, we aimed at a clarification of the concepts and the terminology involved. We reviewed the main features of the approach proposed by Chen \emph{et al.}, where the gauge potential is split into pure-gauge and physical terms, in particular its non-invariance under the so-called Stueckelberg transformations. We showed that Stueckelberg dependence is basically equivalent to background dependence and generalizes the notion of path dependence of non-local gauge-invariant quantities. We also provided an alternative point of view, where Stueckelberg transformations can be understood as combinations of active and passive gauge transformations. We argued that measurability requires gauge invariance but not necessarily Stueckelberg/background/path invariance. The Stueckelberg/background/path dependence being basically fixed by the factorization procedure, we concluded that the Chen \emph{et al.} and Wakamatsu decompositions are perfectly acceptable decompositions.

\section*{Acknowledgments}

In this study, I greatly benefited from numerous discussions with E. Leader and  M. Wakamatsu. This work was supported by the P2I (``Physique des deux Infinis'') network.



\begin{thebibliography}{00}  


\bibitem{Chen:2008ag} 
  X.-S. Chen, X.-F. Lu, W.-M. Sun, F. Wang and T. Goldman,
  {\it Phys. Rev. Lett. } {\bf 100}, 232002 (2008).

\bibitem{Chen:2009mr} 
  X.~-S.~Chen, W.~-M.~Sun, X.~-F.~Lu, F.~Wang and T.~Goldman,
  {\it Phys. Rev. Lett.}  {\bf 103}, 062001 (2009).

\bibitem{Jauch}
  J.~Jauch and F. Rohrlich, 
  {\it The Theory of Photons and Electrons}, 
  (Springer-Verlach, Berlin, 1976).

\bibitem{Berestetskii}
  V.~Berestetskii, E.~Lifshitz, and L.~Pitaevskii, 
  {\it Quantum Electrodynamics}, 
 ( Pergamon, Oxford, 1982).

\bibitem{Jaffe:1989jz} 
  R.~L.~Jaffe and A.~Manohar,
  {\it Nucl. Phys. B} {\bf 337}, 509 (1990).

\bibitem{Ji:1996ek} 
  X.~-D.~Ji,
  {\it Phys. Rev. Lett.}  {\bf 78}, 610 (1997).

\bibitem{Lorce:2012rr} 
  C.~Lorc\'e,
  {\it Phys. Rev. D} {\bf 87}, 034031 (2013).

\bibitem{Wakamatsu:2010qj} 
  M.~Wakamatsu,
  {\it Phys. Rev. D} {\bf 81}, 114010 (2010).

\bibitem{Wakamatsu:2010cb} 
  M.~Wakamatsu,
  {\it Phys. Rev. D} {\bf 83}, 014012 (2011).

\bibitem{Chen:2012vg} 
  X.~-S.~Chen,
  arXiv:1203.1288 [hep-ph].

\bibitem{Lorce:2013gxa} 
  C.~Lorc\'e,
  arXiv:1302.5515 [hep-ph].

\bibitem{Dirac:1955uv} 
  P.~A.~M.~Dirac,
  {\it Can. J. Phys.}  {\bf 33}, 650 (1955).

\bibitem{DeWitt:1962mg} 
  B.~S.~DeWitt,
  {\it Phys. Rev.}  {\bf 125}, 2189 (1962).

\bibitem{Mandelstam:1962mi} 
  S.~Mandelstam,
  {\it Annals Phys.}  {\bf 19}, 1 (1962).

\bibitem{Stoilov:2010pv} 
  M.~N.~Stoilov,
  arXiv:1011.5617 [hep-th].

\bibitem{Zhang:2011rn} 
  P.~M.~Zhang and D.~G.~Pak,
  {\it Eur. Phys. J. A} {\bf 48}, 91 (2012).

\bibitem{Wakamatsu:2013voa} 
  M.~Wakamatsu,
  arXiv:1302.5152 [hep-ph].

\bibitem{Smolin:2005mq} 
  L.~Smolin,
  In {\it The structural foundations of quantum gravity}, eds. D. Rickles \emph{et al.} (Oxford, 2007), p. 196. 

\bibitem{Rickles} 
  D.~Rickles,
  In {\it The Ontology of Spacetime II}, ed. D. Dieks (Elsevier, 2008), p. 133.

\bibitem{Rozali:2008ex} 
  M.~Rozali,
  {\it Adv. Sci. Lett.}  {\bf 2}, 244 (2009).

\bibitem{Barenz:2012av} 
  M.~B\"arenz,
  arXiv:1207.0340 [gr-qc].

\bibitem{Lorce:2013bja} 
  C.~Lorc\'e,
  arXiv:1306.0456 [hep-ph].

\bibitem{Hatta:2011zs} 
  Y.~Hatta,
  {\it Phys. Rev. D} {\bf 84}, 041701 (2011).

\bibitem{Hatta:2011ku} 
  Y.~Hatta,
  {\it Phys. Lett. B} {\bf 708}, 186 (2012).

\bibitem{Hoodbhoy:1998bt} 
  P.~Hoodbhoy, X.~-D.~Ji and W.~Lu,
  {\it Phys. Rev. D} {\bf 59}, 074010 (1999).

\bibitem{Ji:2012sj} 
  X.~Ji, X.~Xiong and F.~Yuan,
  {\it Phys. Rev. Lett.}  {\bf 109}, 152005 (2012).

\bibitem{Manohar:1990kr} 
  A.~V.~Manohar,
  {\it Phys. Rev. Lett.}  {\bf 65}, 2511 (1990).

\bibitem{Guo:2012wv} 
  Z.~-Q.~Guo and I.~Schmidt,
  arXiv:1210.2263 [hep-ph].

\bibitem{Guo:2013jia} 
  Z.~-Q.~Guo and I.~Schmidt,
  arXiv:1303.7210 [hep-ph].

\bibitem{Lorce:2012ce} 
  C.~Lorc\'e,
  {\it Phys. Lett. B} {\bf 719}, 185 (2013).

\bibitem{Collins}
  J.C.~Collins, D.E.~Soper, and G.~Sterman, 
  {\it Perturbative Quantum Chromodynamics}, 
  ed. A.H.~Mueller (World Scientific, Singapore, 1989).



\end{thebibliography}
\end{document}